# New Approaches for Calculating Moran's Index of Spatial Autocorrelation

Yanguang Chen

Department of Geography, College of Urban and Environmental Sciences, Peking University, 100871, Beijing, China. Email: chenyg@pku.edu.cn

**Abstract:** Spatial autocorrelation plays an important role in geographical analysis; however, there is still room for improvement of this method. The formula for Moran's index is complicated, and several basic problems remain to be solved. Therefore, I will reconstruct its mathematical framework using mathematical derivation based on linear algebra and present four simple approaches to calculating Moran's index. Moran's scatterplot will be ameliorated, and new test methods will be proposed. The relationship between the global Moran's index and Geary's coefficient will be discussed from two different vantage points: spatial population and spatial sample. The sphere of applications for both Moran's index and Geary's coefficient will be clarified and defined. One of theoretical findings is that Moran's index is a characteristic parameter of spatial weight matrices, so the selection of weight functions is very significant for autocorrelation analysis of geographical systems. A case study of 29 Chinese cities in 2000 will be employed to validate the innovatory models and methods. This work is a methodological study, which will simplify the process of autocorrelation analysis. The results of this study will lay the foundation for the scaling analysis of spatial autocorrelation.

# 1 Introduction

The theory of spatial autocorrelation has been a key element of geographical analysis for more than twenty years. A number of measurements of spatial correlation were proposed so that we can investigate the spatial process of geographical evolution from differing points of view (Anselin, 1995; Bivand *et al*, 2009; Cliff and Ord, 1969; Cliff and Ord, 1981; Getis and Ord, 1992; Griffith, 2003; Haggett *et al*, 1977; Haining, 2009; Li *et al*, 2007; Odland, 1988; Sokal and Oden, 1978; Sokal and Thomson, 1987; Tiefelsdorf, 2002; Weeks *et al*, 2004). Although there are various correlation measurements, two are commonly used. One is Moran's index (Moran, 1948), and the other, is Geary's coefficient (Geary, 1954). The former is a generalization of Pearson's correlation coefficient, and the latter is analogous to the Durbin-Watson statistic of regression analysis. In theory, Moran's index is somewhat equivalent to Geary's coefficient and they can be substituted for one another. However, in practice, Moran's index cannot be replaced by Geary's coefficient and *vice versa* due to a subtle difference of statistical treatment. Compared with Geary's coefficient, Moran's index is more significant to spatial analysis.

Today, the concepts and methods of spatial autocorrelation have been applied to many fields, which have resulted in a number of interesting findings (Beck and Sieber, 2010; Benedetti-Cecchi *et al*, 2010; Bonnot *et al*, 2010; Braun *et al*, 2012; Chen, 2012; Deblauwe *et al*, 2012; Impoinvil *et al*, 2011; Kumar *et al*, 2012; Mateo-Tomás and Olea, 2010; Stark *et al*, 2012). However, despite its long history, many basic and important questions remain to be answered. For example, we still don't know how to determine the spatial contiguity matrix objectively (Chen, 2012). The relationships between Moran's index and Geary's coefficient are still unclear. In fact, spatial autocorrelation is a special case of the spatial correlation function. This correlation in geographic systems is often associated with the scaling process (Chen, 2009; Chen, 2011; Chen, 2013; Chen and Jiang, 2010). However, so far, spatial autocorrelation has not been linked to scaling laws. It is necessary to find new ways of understanding and implementing spatial autocorrelation analysis in order to solve many theoretical and methodological problems in this field. Before this aim can be accomplished, a simple and general framework must be constructed for this theory.

In this paper, a new way is proposed to express and estimate Moran's index. Geary's coefficient can be re-expressed in a new form along with other related measurements. By doing so, we can further develop the analytical process of spatial autocorrelation. The remaining sections of the paper are organized as follows. In Section 2, I will reconstruct the spatial weight matrix and Moran's index, and improve Moran's scatterplot in terms of the mathematical processes in this study. In Section 3, the association of Moran's index with Geary's coefficient is discussed. The scopes of application of the two parameters are defined for geographical analysis. In Section 4, I will introduce four approaches to calculating Moran's index and a simple approach to computing Geary's coefficient. The principal cities of China, including national capital and provincial capitals, are taken as a case to show how to use the methods advanced in this article. Finally, the paper is concluded with a brief summary. There are several innovative aspects of this study. First, the mathematical expressions are regularized, and new methods of computing spatial autocorrelation measurements are proposed. Second, the implication of Moran's index as a scaling parameter is reviewed. Third, the similarities and differences between Moran's index and Geary's coefficient are clarified. Especially, the threshold values of the two parameters indicating no spatial autocorrelation are corrected for spatial samples.

## 2 Results

### 2.1 Reconstructing Moran's index and spatial weights matrix

The first basic measurement of spatial autocorrelation is Moran's index, which came about as a result of Pearson's correlation coefficient in general statistics (Moran, 1948). Generalizing Pearson's cross-correlation coefficient of two samples to the autocorrelation coefficient of one sample, and then generalizing the 1-dimensional autocorrelation coefficient from time series to the 2-dimensional autocorrelation coefficient about spatial distribution by substituting the weighting function for the lag parameter, we can obtain the formula of Moran's index. Correspondingly, we can derive Geary's coefficient from the idea of spatial autocorrelation by analogy with Watson-Durbin's statistic (Geary, 1954). It will be demonstrated that Moran's index is based on populations, while Geary's coefficient is based on samples. Moran's index is in fact a standardized spatial auto-covariance, which can be simply reinterpreted with linear algebra. Suppose there are $n$ elements (e.g., cities) in a system (e.g., a network of cities) which can be measured by a variable (e.g., city size), $x$. A vector can be defined in the equation below:

$$x = \begin{bmatrix} x_1 & x_2 & \cdots & x_n \end{bmatrix}^\mathrm{T}, \qquad (1)$$

where $x_i$ is a size measurement of the $i$th element ($i$=1,2,…,$n$). The mean of $x_i$ is given in the following equation:

$$\mu = \frac{1}{n}\sum_{i=1}^{n} x_i . \tag{2}$$

The centralized variable can be calculated by

$$y = x - \mu, \tag{3}$$

where $\mu$ represents the average value of a vector consisting of $n$ algebraic/numeric quantities. The population variance is as below:

$$\sigma^2 = \frac{1}{n}\sum_{i=1}^{n}(x_i - \mu)^2 = \frac{1}{n}(x-\mu)^{\mathrm{T}}(x-\mu) = \frac{1}{n}y^{\mathrm{T}}y, \tag{4}$$

where $\sigma$ is the population standard deviation (PSD). The result of scaling transform of the centralized variable forms a standardized vector as follows

$$z = \frac{x-\mu}{\sigma} = \frac{y}{\sigma}, \tag{5}$$

which is termed $z$-score in statistics. It can be shown that the norm of $z$, i.e., the length of the vector, $\|z\|$, exactly equals the dimension of the system, i.e., the number of elements in the system, $n$. Thus we have

$$\|z\| = z^{\mathrm{T}}z = \sum_{i=1}^{n} z_i^2 = \sum_{i=1}^{n}(\frac{x_i-\mu}{\sigma})^2 = \frac{n}{\sigma^2}\frac{1}{n}\sum_{i=1}^{n}(x_i-\mu)^2 = n . \tag{6}$$

Based on the equations prepared above, Moran's index can be reconstructed in a simple way. Suppose there is an $n$-by-$n$ unitary spatial weights matrix (USWM) such as

$$W = [w_{ij}]_{n \times n} . \tag{7}$$

The three properties of the matrix are as follows: (1) Symmetry, i.e., $w_{ij}=w_{ji}$; (2) Zero diagonal elements, namely, $|w_{ii}|=0$, which implies that the entries in the diagonal are all 0; (3) Normalization condition, that is

$$\sum_{i=1}^{n}\sum_{j=1}^{n} w_{ij} = 1 . \tag{8}$$

Then Moran's index can be expressed in the quadratic form:

$$I = z^{\mathrm{T}}Wz, \tag{9}$$

which is simple and more convenient than the conventional expression of Moran's index for the mathematical transform in this context. Expanding equation (9) yields the original formula of Moran's index and provides an autocorrelation coefficient defined in 2-dimensional space

$$I = \frac{y^{\mathrm{T}}(nW)y}{y^{\mathrm{T}}y} = \frac{n\sum_{i=1}^{n}\sum_{j=1}^{n}v_{ij}(x_i-\mu)(x_j-\mu)}{\sum_{i=1}^{n}\sum_{j=1}^{n}v_{ij}\sum_{i=1}^{n}(x_i-\mu)^2}, \quad (10)$$

in which $v_{ij}$ denotes the elements of a spatial contiguity matrix, $V$, which will be defined and discussed in Section 4 (*Materials and Methods*). Equation (10) is the common mathematical form of Moran's index.

The theoretical eigen equation of Moran's index can be derived from the abovementioned definitions. Equation (9) multiplied left by $z$ on both sides of the equal sign yields

$$M^{*}z = zz^{\mathrm{T}}Wz = Iz, \quad (11)$$

where

$$M^{*} = zz^{\mathrm{T}}W \quad (12)$$

can be termed the Ideal Spatial Weights Matrix (ISWM) in a theoretical sense. In equation (11), $z$ is the eigenvector (characteristic vector) of $M^{*}$ and Moran's index is the corresponding maximum eigenvalue (characteristic root) in terms of the absolute value. According to equation (6), normalizing $z$ leads to $z/\sqrt{n}$. It can be proved that the diagonal of $M^{*}$ provides the Local Indicators of Spatial Association (LISA), that is, the local Moran's index defined by Anselin (1995). The entries of $M^{*}$'s diagonal can be generally expressed as

$$I_i = \frac{ny_i\sum_{j=1}^{n}w_{ij}y_j}{y^{\mathrm{T}}y} = z_i\sum_{j=1}^{n}w_{ij}z_j, \quad (13)$$

where $I_i$ refers to the local Moran's index. Accordingly, $I$ denotes the global Moran's index. In fact, due to $w_{ij}=w_{ji}$, for arbitrary $n$, equation (12) can be expanded as follows

$$\begin{bmatrix}z_1\\z_2\\\vdots\\z_n\end{bmatrix}\begin{bmatrix}z_1 & z_2 & \cdots & z_n\end{bmatrix}\begin{bmatrix}w_{11} & w_{12} & \cdots & w_{1n}\\w_{21} & w_{22} & \cdots & w_{2n}\\\vdots & \vdots & \ddots & \vdots\\w_{n1} & w_{n2} & \cdots & w_{nn}\end{bmatrix} = \begin{bmatrix}z_1\sum_{j=1}^{n}w_{1j}z_j & z_1\sum_{j=1}^{n}w_{2j}z_j & \cdots & z_1\sum_{j=1}^{n}w_{nj}z_j\\z_2\sum_{j=1}^{n}w_{1j}z_j & z_2\sum_{j=1}^{n}w_{2j}z_j & \cdots & z_2\sum_{j=1}^{n}w_{nj}z_j\\\vdots & \vdots & \ddots & \vdots\\z_n\sum_{j=1}^{n}w_{1j}z_j & z_n\sum_{j=1}^{n}w_{2j}z_j & \cdots & z_n\sum_{j=1}^{n}w_{nj}z_j\end{bmatrix}.$$

(14)

Comparing equation (14) with equation (13) shows that the elements in the diagonal of $M^{*}$ give the local Moran's index. The trace of $M^*$ is just equal to the global Moran's index.

## 2.2 Actual spatial weights matrix

The practical spatial weights matrix is different from the ISWM. In practice of spatial analysis, the matrix $zz^T$ in equation (11) can be replaced by $z^Tz=n$. It can be shown that $n$ is the maximum eigenvalue of the matrix $zz^T$, and $z$ is the corresponding eigenvector. Considering equation (6), for arbitrary $n$, we have

$$zz^Tz = zn = nz = z^Tzz. \tag{15}$$

Expanding equation (15) yields

$$\begin{bmatrix} z_1 \\ z_2 \\ \vdots \\ z_n \end{bmatrix} \begin{bmatrix} z_1 & z_2 & \cdots & z_n \end{bmatrix} \begin{bmatrix} z_1 \\ z_2 \\ \vdots \\ z_n \end{bmatrix} = \begin{bmatrix} z_1 \sum_{i=1}^n z_i^2 \\ z_2 \sum_{i=1}^n z_i^2 \\ \vdots \\ z_n \sum_{i=1}^n z_i^2 \end{bmatrix} = n \begin{bmatrix} z_1 \\ z_2 \\ \vdots \\ z_n \end{bmatrix}. \tag{16}$$

For example, for $n=2$, the extended form of equation (15) is

$$\begin{bmatrix} z_1 \\ z_2 \end{bmatrix} \begin{bmatrix} z_1 & z_2 \end{bmatrix} \begin{bmatrix} z_1 \\ z_2 \end{bmatrix} = \begin{bmatrix} z_1z_1 & z_1z_2 \\ z_2z_1 & z_2z_2 \end{bmatrix} \begin{bmatrix} z_1 \\ z_2 \end{bmatrix} = \begin{bmatrix} z_1(z_1^2+z_2^2) \\ z_2(z_1^2+z_2^2) \end{bmatrix} = 2\begin{bmatrix} z_1 \\ z_2 \end{bmatrix}. \tag{17}$$

This illustrates that $n$ is one of the eigenvalues of the matrix $zz^T$ corresponding to the eigenvector $z$. Further, it can be shown that $n$ is the maximum eigenvalue of $zz^T$. For a square matrix, the trace of $zz^T$ is

$$\mathrm{T_r}(zz^T) = z_1^2 + z_2^2 + \cdots z_n^2 = n = \lambda_1 + \lambda_2 + \cdots + \lambda_n, \tag{18}$$

where $\mathrm{T_r}$ refers to "finding the trace (of $zz^T$)". If $\lambda_1=\lambda_{max}=n$ as given, then

$$\lambda = \begin{cases} n, & \lambda = \lambda_{max} \\ 0, & \lambda \neq \lambda_{max} \end{cases}. \tag{19}$$

For arbitrary $n$, the extended form of $zz^T$ is below

$$zz^T = \begin{bmatrix} z_1 \\ z_2 \\ \vdots \\ z_n \end{bmatrix} \begin{bmatrix} z_1 & z_2 & \cdots & z_n \end{bmatrix} = \begin{bmatrix} z_1z_1 & z_1z_2 & \cdots & z_1z_n \\ z_2z_1 & z_2z_2 & \cdots & z_2z_n \\ \vdots & \vdots & \ddots & \vdots \\ z_nz_1 & z_nz_1 & \cdots & z_nz_n \end{bmatrix}. \tag{20}$$

According to the Cayley-Hamilton theorem, the eigenvalues of any $n$-by-$n$ matrix are identical to the roots of a polynomial equation. For example, for $n=2$, the characteristic polynomial of the matrix $zz^T$ is

$$\lambda E - zz^T = \begin{vmatrix} \lambda - z_1 z_1 & -z_1 z_2 \\ -z_1 z_2 & \lambda - z_2 z_2 \end{vmatrix} = \lambda^2 - \lambda(z_1^2 + z_2^2) = \lambda^2 - 2\lambda = 0, \tag{21}$$

where $E$ denotes the identity/unit matrix. Thus

$$\lambda_1 = \sum_{i=1}^{2} z_i^2 = 2, \quad \lambda_2 = 0. \tag{22}$$

The conclusion can be drawn that the maximum eigenvalue of matrix $zz^T$ is its dimension.

Substituting the maximum eigenvalue $n$ for the corresponding matrix $zz^T$ in equation (11) will provide a new mathematical relationship. In fact, the precondition that equation (10) comes into existence is as below

$$nWy = Iy. \tag{23}$$

In other words, from equation (23) it follows equation (10). Apparently, Moran's index is the maximum eigenvalue of $nW$, and $y$ is the corresponding eigenvector, which can be normalized as $z/\sqrt{n}$. Equation (23) divided by the standard error $\sigma$ on both sides yields

$$nW\frac{y}{\sigma} = I\frac{y}{\sigma}. \tag{24}$$

This leads to the following scaling relationship:

$$Mz = nWz = Iz, \tag{25}$$

where

$$M = nW = z^T zW. \tag{26}$$

This is the Real Spatial Weights Matrix (RSWM) in the sense of application or practice. What is referred to as the "spatial weights matrix" in the literature is just RSWM rather than $W$ or ISWM. The trace of the matrix $nW$ is the eigenvalue with the minimum absolute value, i.e. $T_r(nW)=0$. Normalizing the eigenvector yields

$$z° = \frac{z}{\sqrt{\|z\|}} = \frac{z}{\sqrt{n}}. \tag{27}$$

If we employ mathematical software such as Matlab and Mathcad to calculate the eigenveactor of $zz^T W$ or $nW$, the result will be $z°$ instead of $z$. Comparing equation (25) with equation (11) shows

$$zz^TWz = nWz. \tag{28}$$

This suggests that when the eigenvector $z$ is multiplied by $W$ on the left side, it will remain the eigenvector of $zz^T$. Thus we have

$$(nE - zz^T)Wz = (nW - zz^TW)z = 0, \tag{29}$$

in which 0 refers to the zero/null vector. However, equation (29) cannot occur unconditionally. In empirical analysis, the null vector should be replaced by a residual vector. An approximation relation is as follows

$$Mz = nWz \rightarrow zz^TWz = M^*z, \tag{30}$$

where the arrow "→" denotes "infinitely approach to" or "be theoretically equal to". Empirically, there are always errors between $M=z^TzW$ and $M^*=zz^TW$, which will lead to a new approach to testing the spatial autocorrelation analysis. If the spatial autocorrelation is very strong, $Mz$ will be a very close approximation to $M^*z$; otherwise, the two vectors will be significantly different.

The concept of invariance in the process of mathematical transform is very important for geographical modeling. Equation (11) and Equation (25) are two eigenequations that suggest some invariance in the mathematical transform. The invariance in a transform suggests some invariance behind change or some robustness in the process of spatio-temporal evolution of a system. A characteristic equation denotes a scaling relationship of matrices. The invariance in change or transform implies some kind of symmetry, which indicates some law of conservation. In geography, symmetry is an essential criterion for model building, method choice, and parameter estimation (Chen, 2013). Moran's index is a quantity of invariance in mathematical transform, so it is a very basic and significant parameter for spatial analysis.

## 2.3 Improved version of Moran's scatterplot

The analytical process of spatial autocorrelation can be developed using the mathematical expressions proposed above. In order to find new approaches to evaluating Moran's index and improve the Moran scatterplot, two vectors based on spatial weights matrix are defined as below

$$f^* = M^*z = zz^TWz = Iz, \tag{31}$$

$$f = Mz = nWz = z^TzWz, \tag{32}$$

where the relationship between $z$ and $f^*$ suggests the theoretical autocorrelation trend, i.e., the regression line, and the dataset of $z$ and $f$, denotes the actual autocorrelation pattern, i.e. the points on the scatter diagram. The residuals of spatial autocorrelation can be defined as

$$e_f = f - f^* = Mz - M^*z = (nE - zz^T)Wz, \tag{33}$$

where $e_f$ refers to the errors of the spatial autocorrelation. The squared sum of the residuals $S_f$ is

$$S_f = e_f^T e_f = z^T W(nE - zz^T E)(nE - zz^T)Wz \to 0. \tag{34}$$

The value of $e_f$ fluctuates around 0; therefore, the $S_f$ value approaches zero. A standard error can be defined in the form

$$s_f = \sqrt{\frac{1}{n}S_f} = \sqrt{\frac{1}{n}e_f^T e_f}. \tag{35}$$

in which $s_f$ refers to the standard error between the variables $f$ and $f^*$.

The error sum of square can be equivalently expressed in another form. From equation (11) it follows the observed values of $z$ as below:

$$z = \frac{1}{I}zz^T Wz = \frac{1}{I}f^*. \tag{36}$$

Correspondingly, the predicted values of $z$ can be given by

$$z^* = \frac{1}{I}nWz = \frac{1}{I}f. \tag{37}$$

Thus another residual vector is as below:

$$e_z = z - z^* = \frac{1}{I}(zz^T - nE)Wz = -\frac{1}{I}e_f, \tag{38}$$

where $e_z$ denotes the residual vector of $z$ and $z^*$. Obviously, the two residual vectors, $e_z$ and $e_f$, are equivalent to one another. The squared sum of the residuals $e_z$ is

$$S_z = e_z^T e_z = \frac{1}{I^2}e_f^T e_f \to 0. \tag{39}$$

Accordingly, another standard error can be defined as follows

$$s_z = \sqrt{\frac{1}{n}S_z} = -\frac{s_f}{I}. \tag{40}$$

in which $s_z$ refers to the standard error between the variables $z$ and $z^*$. This implies that the two standard errors, $s_f$ and $s_z$, are equivalent to each other.

After evaluating Moran's index, it can be tested in two ways. First, the series of residuals should satisfy the normal distribution, which shows a bell-shaped curve or histogram. If this distribution does not occur, the spatial weights matrix must be adjusted or the weight function must be changed. Second, the standard error between $f$ and $f^*$ should be less than 0.15, that is, $s_f < 0.15$, which is an empirical value. The standard errors and the histograms of residuals make two test approaches of spatial autocorrelation.

Moran's scatterplot can be amended and developed using the above equations. The Moran scatterplot proposed by Anselin (1996) plays an important part in spatial autocorrelation analysis. If $y$ represents the $x$-axis, and $nWy$ represents the $y$-axis, a conventional Moran scatterplot can be created. The scatterplot can be improved as follows. First, a trend line can be added to the plot. Second, the variables can be standardized, and $x$ or $y$ can be replaced by $z$. Based on equations (31) and (32), the Moran scatterplot can be bettered so that it will illustrate spatial autocorrelation more efficiently. The plot of $f^*$ vs. $z$ shows a set of ordered data points, which make a straight line, while the plot of $f$ vs. $z$ displays a set of randomly scattered data points. Superimposing the two plots onto each other yields an improved scatter diagram for Moran's $I$. In the revised plot, the coordinates $(z_i, f_i^*)$ represent the ideal locations that form a trend line, while $(z_i, f_i)$ represent the actual locations of data points that are irregularly scattered. The slopes of the trend lines of $(z_i, f_i^*)$ and $(z_i, f_i)$ indicate Moran's index. Thus, the geometric meaning of the Moran scatterplot becomes clear. Moreover, an inverse Moran scatterplot can be defined based on equations (36) and (37). In the inverse scatterplot, the abscissa ($x$-axis) is represented by $f^*$, and the ordinate ($y$-axis) is represented by $z$ and $z^*$. The coordinates $(f_i^*, z_i)$ indicates the ideal locations which form a straight line, while $(f_i^*, z_i^*)$ denotes the actual locations of data points which are irregularly scattered. A dual relationship can be found between the Moran scatterplot and the inverse Moran scatterplot.

## 3 Discussion

### 3.1 Revision of Moran's index and Geary's coefficient

Moran's index is only one of the many spatial autocorrelation measurements in geographical analysis. Another important measurement is Geary's coefficient (Geary, 1954). In theory, Moran's index can be associated with Geary's coefficient. However, the former cannot be directly converted into the latter because that the bases of definitions of the two spatial statistics are different. Geary's coefficient is defined in the equation below:

$$C = \frac{(n-1)\sum_{i=1}^{n}\sum_{j=1}^{n}v_{ij}(x_i - x_j)^2}{2\sum_{i=1}^{n}\sum_{j=1}^{n}v_{ij}\sum_{i=1}^{n}(x_i - \bar{x})^2} = \frac{(n-1)\sum_{i=1}^{n}\sum_{j=1}^{n}w_{ij}(x_i - x_j)^2}{2\sum_{i=1}^{n}(x_i - \bar{x})^2}, \tag{41}$$

in which $\bar{x} = \mu$ is the mean of $x$, and $v_{ij}$ is the spatial contiguity measurement (See Subsection 4.1). Equation (41) can be rewritten as

$$C = \frac{1}{2}\sum_{i=1}^{n}\sum_{j=1}^{n}w_{ij}(\frac{x_i - \bar{x}}{s} - \frac{x_j - \bar{x}}{s})^2 = \frac{1}{2}\sum_{i=1}^{n}\sum_{j=1}^{n}w_{ij}(Z_i - Z_j)^2, \tag{42}$$

where $s$ refers to a sample standard deviation (SSD), and $Z$, to the corresponding standardized vector, that is

$$s = \sqrt{\frac{1}{n-1}\sum_{i=1}^{n}(x_i - \bar{x})^2}, Z = \frac{x - \bar{x}}{s}.$$

This differs from the definition of Moran's index, which, as indicated above, is based on PSD. The traditional Moran's index is used to describe spatial population, while traditional Geary's coefficient is for describing a spatial sample.

The base of spatial analysis can be divided into two cases: spatial population and spatial sample. Based on spatial sample, Moran's index can be revised as

$$I^* = \frac{(n-1)\sum_{i=1}^{n}\sum_{j=1}^{n}v_{ij}(x_i - \bar{x})(x_j - \bar{x})}{\sum_{i=1}^{n}\sum_{j=1}^{n}v_{ij}\sum_{i=1}^{n}(x_i - \bar{x})^2} = Z^T W Z = \frac{n-1}{n}z^T W z = \frac{n-1}{n}I, \tag{43}$$

which suggests that there is a linear scaling relationship between the population-based Moran's index and the sample-based Moran's index. On the other hand, based on a spatial population, Geary's coefficient can be redefined as

$$C^* = \frac{n\sum_{i=1}^{n}\sum_{j=1}^{n}v_{ij}(x_i - x_j)^2}{2\sum_{i=1}^{n}\sum_{j=1}^{n}v_{ij}\sum_{i=1}^{n}(x_i - \mu)^2} = \frac{1}{2}\sum_{i=1}^{n}\sum_{j=1}^{n}w_{ij}(z_i - z_j)^2 = \frac{n}{n-1}C. \tag{44}$$

Rearranging equation (44) yields

$$C^* = \sum_{i=1}^{n}\sum_{j=1}^{n}w_{ij}(z_i^2 - z_i z_j) = \sum_{i=1}^{n}\sum_{j=1}^{n}w_{ij}z_i^2 - z^T W z. \tag{45}$$

Defining a parameter such as

$$\omega = \sum_{i=1}^{n}\sum_{j=1}^{n} w_{ij} z_i^2 \to \frac{1}{n}\sum_{i=1}^{n} z_i^2 = 1, \quad (46)$$

we have

$$C^* = \omega - I \to 1 - I, \quad (47)$$

which reflects the relation between Moran's index and the Geary's coefficient based on spatial population. Since $I$ ranges from -1 to 1, $C^*$ will have a value between 0 and 2. The threshold value of Geary's coefficient is $C_t^*=1$, where the subscript "t" refers to "threshold"; therefore, the threshold value of Moran's index is $I_t=0$, indicating no spatial autocorrelation. In light of equation (47), if $C^*<1$, then $I>0$, and then there will be a positive spatial autocorrelation (PSAC); If $C^*>1$, then $I<0$, then there will be a negative spatial autocorrelation (NSAC).

Similarly, for a spatial sample, the mathematical relationship between the revised Moran's index and Geary's coefficient can be derived as follows:

$$C = \psi - I^* \to \frac{n-1}{n} - I^*, \quad (48)$$

where

$$\psi = \sum_{i=1}^{n}\sum_{j=1}^{n} w_{ij} Z_i^2 \to \frac{1}{n}\sum_{i=1}^{n} Z_i^2 = \frac{n-1}{n}. \quad (49)$$

This suggests that, for a spatial sample, the threshold value of Geary's coefficient is $C_t=(n-1)/n$, which implies no autocorrelation in a spatial process. If $C_t<(n-1)/n$, then $I^*>0$, there will be a PSAC; If $C_t>(n-1)/n$, then $I^*<0$, and there will be a NSAC.

## 3.2 Thresholds of Moran's index and Geary's coefficient

A long-standing problem about the threshold values of Moran's index and Geary's coefficient, which suggest no spatial autocorrelation, should be solved. For a spatial population, the critical value of Moran's index is $I_t=0$, and the threshold of Geary's coefficient is $C_t^*=1$. These are indisputable. However, for a spatial sample, the consensus has not yet been reached so far. In many previous works, the threshold value of Moran's index was regarded as $I_t^*=1/(1-n)$ (Odland, 1988). If this were true, then, according to equation (43), the threshold of Moran's index for population would be $I_t=-n/(n-1)^2$; According to equation (47), the threshold of Geary's coefficient for population would be $C_t^*=[(n-1)^2+n]/(n-1)^2$; According to equation (44) or equation (48), the threshold of Geary's coefficient for sample would be $C_t=[(n-1)^2+n]/[n(n-1)]$. However, based on Pearson's correlation coefficient, Moran's index will indicate null autocorrelation if and only if its value equals zero. In other words, for populations, the threshold of Moran's index must be $I_t=0$ rather than other values. Then, according to equation (43), we have $I_t^*=0$ for samples; according to equation (47), we have $C_t^*=1$ for populations; according to equation (44) or equation (48), we have $C_t=(n-1)/n$ for samples. A comparison between the new values and the traditional results of autocorrelation thresholds can be drawn as follows (Table 1).

Table 1 The threshold values of Moran's index and Geary's coefficient and the revised results

| Parameter | In this paper (new results) | | In previous literature (old results) | |
| --- | --- | --- | --- | --- |
| | Spatial population | Spatial sample | Spatial population | Spatial sample |
| Moran's index | 0 | 0 | 0 | $-1/(n-1)$ |
| Geary's coefficient | 1 | $(n-1)/n$ | 1 | 1 |

## 4 Materials and Methods

### 4.1 Four approaches to Moran's index

The traditional method of evaluating Moran's index is so complicated that it is difficult for learners to make spatial analyses using the autocorrelation measurement. Based on the new framework of spatial autocorrelation expressed through linear algebra, especially, equations (11), (25), (31), and (32), four simple approaches to computing Moran's index are proposed as follows. The first is a three-step calculation method, the second is the matrix scaling method, the third is the linear regression method, and the fourth is the standard deviation method. The four approaches are theoretically equivalent to one another. However, in practice, each method has its own advantages and disadvantages (Table 2).

Table 2 Comparison of the advantages and disadvantages of the four methods

| Method | Simplicity | Global Moran's $I$ | Local Moran's $I$ | Test |
| --- | --- | --- | --- | --- |
| Three-step calculation | Very simple | Directly yield | Indirectly yield | Complicated |
| Matrix scaling | Simple | Directly yield | Directly yield | Complicated |
| Linear regression | Moderate | Directly yield | Indirectly yield | Simple |
| Standard deviation | Moderate | Directly yield | Indirectly yield | Complicated |

**(1) Three-step calculation method.** This is the basic method, which can be readily mastered by the beginners of spatial autocorrelation analysis. Based on the standardized vector $z$ and the spatial weights matrix $W$, the three steps of calculating Moran's index are as follows. Step 1: standardize the variable. In other words, convert the initial vector in equation (1) into the standardized vector in equation (5). According to the original definition of Moran's index (Moran, 1948), the PSD rather than the SSD should be used to standardize the data. Step 2: calculate the USWM. The weights matrix is defined in equations (7) and (8). Step 3: compute Moran's index. According to equation (9), the USWM is first left multiplied by the transposition of $z$, and then the product of $z^T$ and $W$ is right multiplied by $z$. The final product of the continued multiplication is the value of Moran's index.

**(2) Matrix scaling method.** It can also be termed "characteristic value method". The key is to find the maximum eigenvalue of the matrix $M^*=zz^TW$ or $M=nW$ by using equation (11) or equation (25). In fact, if $M^*$ is obtained through equation (12), the local Moran's index can also be calculated. The values of the principal diagonal elements are just the local Moran's indexes. The trace of $M^*$ is actually the global Moran's index which can be determined by the equation below:

$$I = T_r(M^*) = T_r(zz^TW), \tag{50}$$

where $T_r$ denotes finding the sum of the numbers in the principal diagonal of a matrix.

**(3) Regression analysis method.** Linear regression can be employed to solve equations (31) and (32) and evaluate Moran's index. Suppose that $z$ acts as an independent variable (i.e., argument), and $f^*=M^*z$ or $f=Mz$ as the corresponding dependent variable (response variable). A regression analysis can be conducted by letting the constant equal zero. The regression coefficient (slope) gives the value of Moran's index.

**(4) Standard deviation method.** It can be proved that the PSD of $f^*$ is just the absolute value of Moran's index. In terms of equation (31), the average value of $f^*$ is zero since the mean of $z$ is zero. Considering equation (6), the population variation of $f^*$ is

$$\frac{1}{n}f^{*T}f^* = \frac{1}{n}z^TzI^2 = I^2. \tag{51}$$

Thus the value of Moran's index can be given by

$$I = \pm\sqrt{\frac{1}{n} f^{*T} f^*}. \tag{52}$$

This suggests an alternative approach to estimating Moran's index.

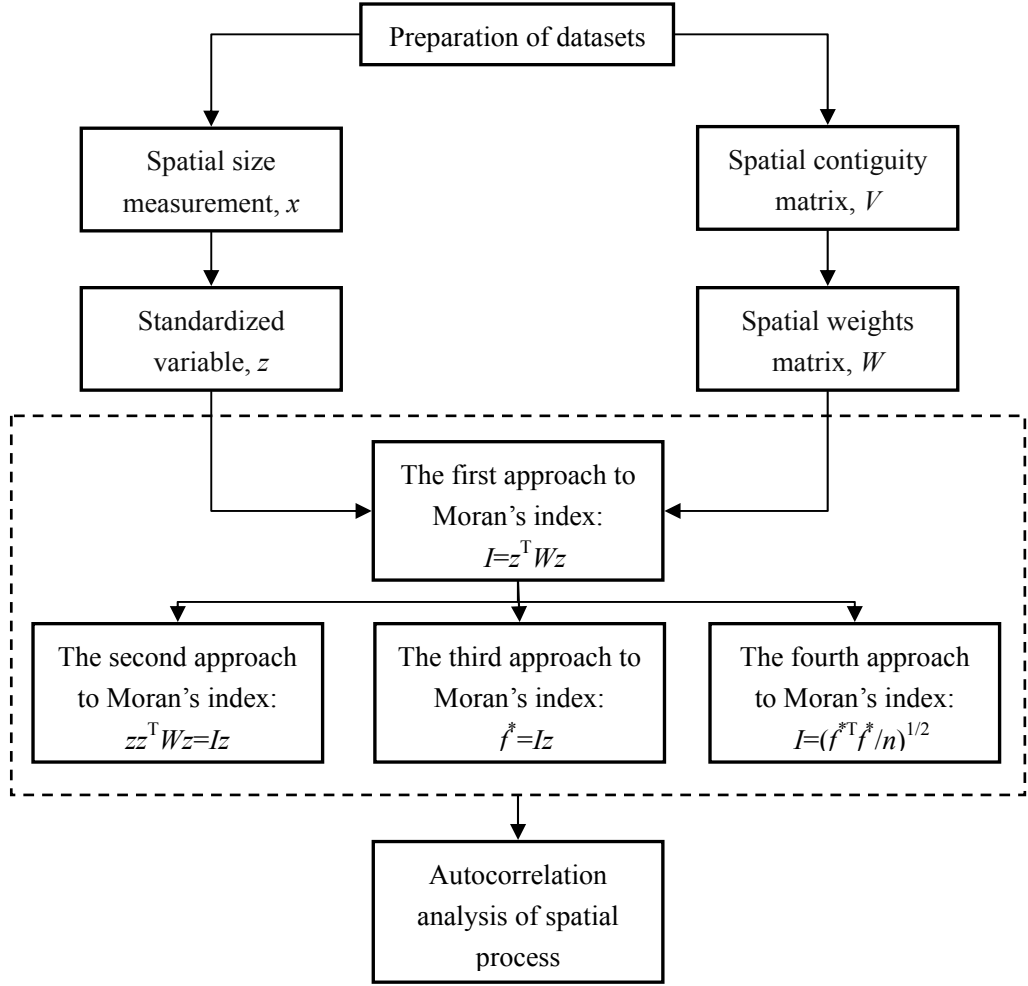

**Figure 1** A flow chart of data processing, parameter estimation, and autocorrelation analysis

The first method is the most important one in this framework of spatial autocorrelation analysis. The second, third, and fourth ones are in fact derived from the first method. The process of data preparation, parameter estimation, and analysis of results based on Moran's index can be illustrated with a flow chart (Figure 1). As an alternative measurement of Moran's index, Geary's coefficient can be utilized to make a spatial analysis. A new approach comprising five steps to computing Geary's coefficient has been found by analogy with the first approach to evaluating Moran's index. However, compared with determining Moran's index, calculating Geary's coefficient is complex to some extent. The first step is to standardize data using the formula $z^{\Delta}=(x-\mu)/s$, where $s$ denotes SSD. The second step is to the compute the squares of difference using the formula $Z_{ij} =(z_i^{\Delta}- z_j^{\Delta})^2$. The results compose a matrix $Z=[(z_i^{\Delta}- z_j^{\Delta})^2]=[Z_{ij}]$. The third step is to transform the spatial contiguity matrix ($V$) into the spatial weights matrix ($W=V/W_0$). The fourth step is to calculate the sum of the products of the algebraic quantities of $W$ and the numeric quantities of $Z$ using the following formula $S=\sum\sum w_{ij}Z_{ij}$. The fifth step is to evaluate Geary's coefficient using the formula $C=S/2$.

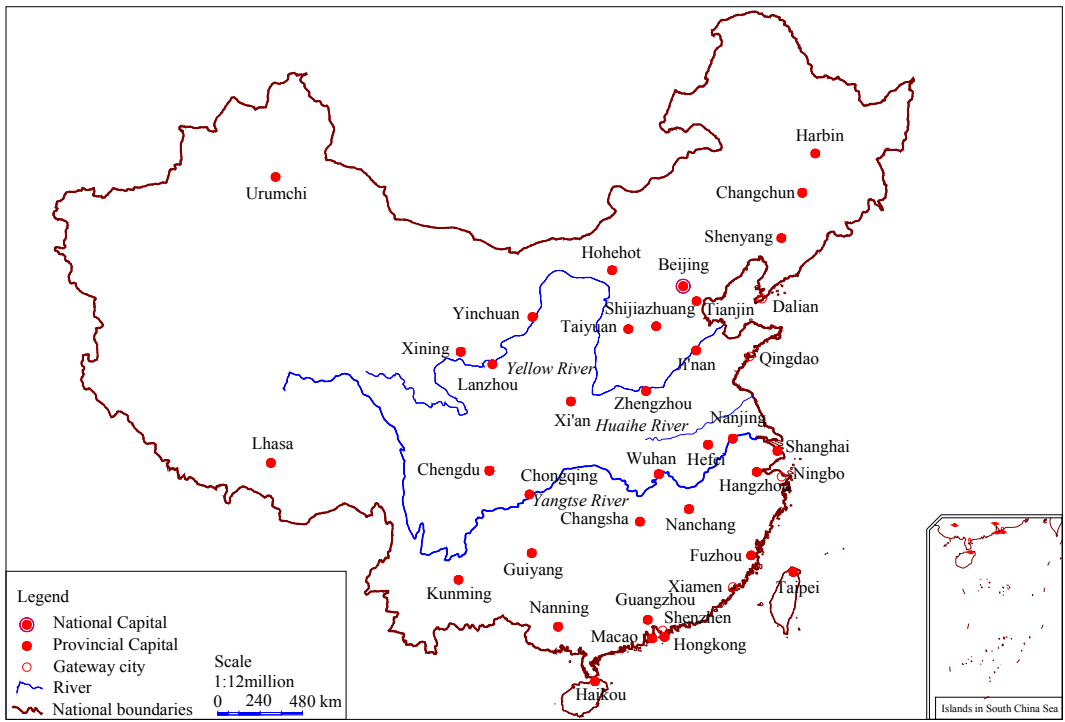

**Figure 2** A sketch map of the geographic relationship of the 31 principal cities of China

## 4.2 Empirical analysis

The improved framework of spatial autocorrelation can be applied to China's cities to make an example of methodological practice. For simplicity, only the capital cities of the 31 provinces, autonomous regions, and municipalities directly under the Central Government of China are considered in this case study (Figure 2). The urban population is employed as a size measurement, while the distances by train between any two cities is used as a spatial contiguity measurement. The census data of the urban population in 2000 are available from the Chinese website (http://pdfdown.edu.cnki.net), and the railroad distance matrix can be found in many Chinese road atlases. Because the two cities of Haikou and Lhasa are not connected to the network of Chinese cities by railway in 2000, 29 cities are actually included in the dataset. In other words, the size of the spatial sample is $n=29$ (Table 2).

In order to construct a spatial weights matrix, a spatial contiguity matrix must be created by using a weight function (Chen, 2012; Getis, 2009). For $n$ elements in a geographic system, a spatial contiguity matrix, $V$, can be expressed as

$$V = [v_{ij}]_{n \times n} = \begin{bmatrix} v_{11} & v_{12} & \cdots & v_{1n} \\ v_{21} & v_{22} & \cdots & v_{2n} \\ \vdots & \vdots & \ddots & \vdots \\ v_{n1} & v_{n2} & \cdots & v_{nn} \end{bmatrix}, \quad (53)$$

in which $v_{ij}$ is a measure used to compare and judge the degree of nearness or the contiguous relationships between location $i$ and location $j$ ($i, j=1,2,\ldots,n$). No matter what the entry $v_{ii}$ equals, it will be converted into zero (for $i=j$, $v_{ii}\equiv 0$). Thus a spatial weights matrix, $W$, can be given by

$$W = \frac{V}{V_0} = \begin{bmatrix} w_{11} & w_{12} & \cdots & w_{1n} \\ w_{21} & w_{22} & \cdots & w_{2n} \\ \vdots & \vdots & \ddots & \vdots \\ w_{n1} & w_{n2} & \ddots & w_{nn} \end{bmatrix} \text{ or } W_* = \frac{nV}{V_0} = n \begin{bmatrix} w_{11} & w_{12} & \cdots & w_{1n} \\ w_{21} & w_{22} & \cdots & w_{2n} \\ \vdots & \vdots & \ddots & \vdots \\ w_{n1} & w_{n2} & \ddots & w_{nn} \end{bmatrix}, \quad (54)$$

where

$$V_0 = \sum_{i=1}^{n}\sum_{j=1}^{n} v_{ij}, \quad w_{ij} = \frac{v_{ij}}{\sum_{i=1}^{n}\sum_{j=1}^{n} v_{ij}}, \quad \sum_{i=1}^{n}\sum_{j=1}^{n} w_{ij} = 1.$$

The value $v_{ii}\equiv 0$ results in the value $w_{ii}\equiv 0$. It is clear that $W$ is mathematically equivalent to $W_*$. In the literature, $W_*$ is often used to represent the spatial weights matrix. However, I employ $W$ to make an empirical analysis of China's cities because the models and methods presented in Section 2 (*Results*) are based on $W$ instead of $W_*$. Compared with $W_*$, $W$ make the mathematical process of spatial autocorrelation become simple and graceful.

Three types of functions can be used as a spatial weight function: inverse power function, negative exponential function, and staircase functions (Chen, 2012). In this case, two functions will be adopted. The first is the inverse power function, which indicates action at a distance or global correlation in geography. Generally, the inverse power function is in the form below:

$$v_{ij} = \begin{cases} r_{ij}^{-b}, & i \neq j \\ 0, & i = j \end{cases}, \quad (55)$$

where $r_{ij}$ refers to the distance between location $i$ and location $j$, and $b$ denotes the distance decay coefficient (generally, $b=1$). This kind of weight function comes from the impedance function of the gravity model (Haggett *et al*, 1977). Cliff and Ord (1973, 1981) used this function to construct the spatial congruity matrix. The second function is the negative exponential function indicative of quasi-global correlation or even quasi-local correlation (Chen, 2008). The exponential function can be expressed as

$$v_{ij} = \begin{cases} \exp(-r_{ij}/\bar{r}), & i \neq j \\ 0, & i = j \end{cases}, \quad (56)$$

where $\bar{r}$ denotes the average distance between any two locations, and it can be defined as the arithmetic mean of all the numbers in a distance matrix. The exponential function can be derived from the entropy- maximizing model proposed by Wilson (1970).

The four approaches discussed in Subsection 4.1 can be employed to estimate Moran's index of the 29 Chinese cities. The process of the basic computation is described below. Using equations (1) - (5), we can standardize the vector of the urban population sizes of the 29 cities $x$ and yield $z$. Then, applying equations (55) or (56) to the matrix of railway distances between the 29 cities yields a spatial contiguity matrix, indicated by equation (53). Let the entries of diagonals, $v_{ii}$, equal zero ($v_{ii}=0$). By referring to the method in which equation (53) transforms into equation (54), we can convert the spatial contiguity matrices ($V$) into the spatial weight matrices ($W$) based on railway distance. Finally, equation (9), or (11), or (31) is utilized to calculate Moran's index. The local Moran index, LISA, can be determined by equation (12) or (13). The diagonal elements of $M^*=zz^TW$ are simply the value of LISA. The main results are displayed in Table 3.

Table 3 Classification of spatial autocorrelation based on population size of the principal cities in China (2000)

| City | $z$ | Based on inverse power function | | | | | Based on negative exponential function | | | | |
|---|---|---|---|---|---|---|---|---|---|---|---|
| | | $f$ | $f^*$ | $z^*$ | LISA | Type | $f$ | $f^*$ | $z^*$ | LISA | Type |
| Beijing | 2.2876 | 0.0236 | -0.0720 | -0.7510 | 0.0019 | H-H | -0.0784 | -0.0939 | 1.9105 | -0.0062 | H-L |
| Changchun | -0.2685 | 0.1215 | 0.0084 | -19.1155 | -0.0011 | L-H | 0.1015 | 0.0110 | -1.6814 | -0.0009 | L-H |
| Changsha | -0.4840 | 0.1389 | 0.0152 | -9.3401 | -0.0023 | L-H | 0.1061 | 0.0199 | -3.3632 | -0.0018 | L-H |
| Chengdu | 0.1544 | -0.0091 | -0.0049 | -1.0662 | 0.0000 | H-L | -0.0436 | -0.0063 | -1.5679 | -0.0002 | H-L |
| Chongqing | 0.8516 | -0.0761 | -0.0268 | -1.7266 | -0.0022 | H-L | -0.0711 | -0.0350 | -1.2534 | -0.0021 | H-L |
| Fuzhou | -0.5373 | 0.0622 | 0.0169 | -2.0959 | -0.0012 | L-H | 0.1019 | 0.0221 | -1.6559 | -0.0019 | L-H |

| City | | | | | | | | | | |
|---|---|---|---|---|---|---|---|---|---|---|
| Guangzhou | 1.2998 | -0.0291 | -0.0409 | -3.8635 | -0.0013 | H-L | -0.0160 | -0.0533 | -2.4726 | -0.0007 | H-L |
| Guiyang | -0.5943 | 0.0509 | 0.0187 | -0.8618 | -0.0010 | L-H | 0.0420 | 0.0244 | -1.6313 | -0.0009 | L-H |
| Hangzhou | -0.3606 | 0.6607 | 0.0113 | 5.9883 | -0.0082 | L-H | 0.1736 | 0.0148 | 3.0223 | -0.0022 | L-H |
| Harbin | 0.0167 | 0.0271 | -0.0005 | -10.4506 | 0.0000 | H-H | 0.0670 | -0.0007 | -3.2167 | 0.0000 | H-H |
| Hefei | -0.7321 | 0.1499 | 0.0230 | -21.0039 | -0.0038 | L-H | 0.1456 | 0.0300 | -4.2294 | -0.0037 | L-H |
| Hohehot | -0.9095 | 0.0543 | 0.0286 | -4.7648 | -0.0017 | L-H | 0.0514 | 0.0373 | -3.5477 | -0.0016 | L-H |
| Jinan | -0.3100 | 0.1776 | 0.0098 | -1.9766 | -0.0019 | L-H | 0.1360 | 0.0127 | -2.4831 | -0.0015 | L-H |
| Kunming | -0.3503 | -0.0158 | 0.0110 | -4.2615 | 0.0002 | L-L | -0.0087 | 0.0144 | -3.2791 | 0.0001 | L-L |
| Lanzhou | -0.5996 | -0.2605 | 0.0189 | -5.6450 | 0.0054 | L-L | -0.0810 | 0.0246 | -3.3135 | 0.0017 | L-L |
| Nanchang | -0.6544 | 0.1341 | 0.0206 | -3.7769 | -0.0030 | L-H | 0.1346 | 0.0269 | -2.5163 | -0.0030 | L-H |
| Nanjing | 0.0137 | 0.3287 | -0.0004 | 3.0971 | 0.0002 | H-H | 0.1320 | -0.0006 | 0.5454 | 0.0001 | H-H |
| Nanning | -0.7728 | 0.0400 | 0.0243 | -4.4145 | -0.0011 | L-H | 0.0709 | 0.0317 | -2.5859 | -0.0019 | L-H |
| Shanghai | 3.4994 | -0.1884 | -0.1101 | 0.9254 | -0.0227 | H-L | -0.1240 | -0.1436 | 0.3894 | -0.0150 | H-L |
| Shenyang | 0.3512 | 0.0659 | -0.0110 | -1.2712 | 0.0008 | H-H | 0.0680 | -0.0144 | -1.7284 | 0.0008 | H-H |
| Shijiazhuang | -0.5563 | 0.2938 | 0.0175 | 2.4202 | -0.0056 | L-H | 0.1380 | 0.0228 | 1.7320 | -0.0026 | L-H |
| Taiyuan | -0.3279 | 0.0335 | 0.0103 | 0.2891 | -0.0004 | L-H | 0.0644 | 0.0135 | 1.0620 | -0.0007 | L-H |
| Tianjin | 0.7153 | 0.6013 | -0.0225 | -1.6172 | 0.0148 | H-H | 0.0690 | -0.0294 | -1.0227 | 0.0017 | H-H |
| Urumchi | -0.6326 | -0.0232 | 0.0199 | 0.5011 | 0.0005 | L-L | -0.0465 | 0.0260 | 0.2124 | 0.0010 | L-L |
| Wuhan | 1.1835 | -0.0974 | -0.0372 | 2.1288 | -0.0040 | H-L | -0.0224 | -0.0486 | 0.6314 | -0.0009 | H-L |
| Xi'an | 0.0882 | -0.0670 | -0.0028 | 8.2811 | -0.0002 | H-L | -0.0259 | -0.0036 | 1.9734 | -0.0001 | H-L |
| Xining | -0.9608 | -0.1557 | 0.0302 | 4.9494 | 0.0052 | L-L | -0.0551 | 0.0394 | 1.3425 | 0.0018 | L-L |
| Yinchuan | -1.0646 | -0.0789 | 0.0335 | 2.5088 | 0.0029 | L-L | -0.0215 | 0.0437 | 0.5231 | 0.0008 | L-L |
| Zhengzhou | -0.3457 | 0.1188 | 0.0109 | 0.7379 | -0.0014 | L-H | 0.1033 | 0.0142 | 1.1327 | -0.0012 | L-H |

It has been demonstrated that the thresholds of Moran's index for both populations and samples are zero (*Discussion*). Based on the inverse power function, equation (55), the value of Moran's index is $I \approx -0.0315 < 0$, which implies a weak negative autocorrelation; Based on the exponential function, equation (56), the index value is $I \approx -0.0410 < 0$, which also suggests a weak negative autocorrelation between the capital cities. These results are consistent with the conclusion drawn from the error sums of square. Using equations (34) and (35), we can compute the squared sum of the residuals between $f$ and $f^*$ and the corresponding standard errors. Based on the power function, the error sum of square is around $S_f=1.2570$, the standard error is about $s_f=0.2082$; Based on the exponential function, we have $S_f=0.1878$ as well as $s_f=0.0805$. These results show that, for the city sizes of China in 2000, the spatial autocorrelation based on the exponential function (confidence level is greater than 99%) is more significant than the correlation based on the power function (confidence level is greater than 59%). This suggests that China's cities are locally spatial autocorrelation rather than global spatial autocorrelation.

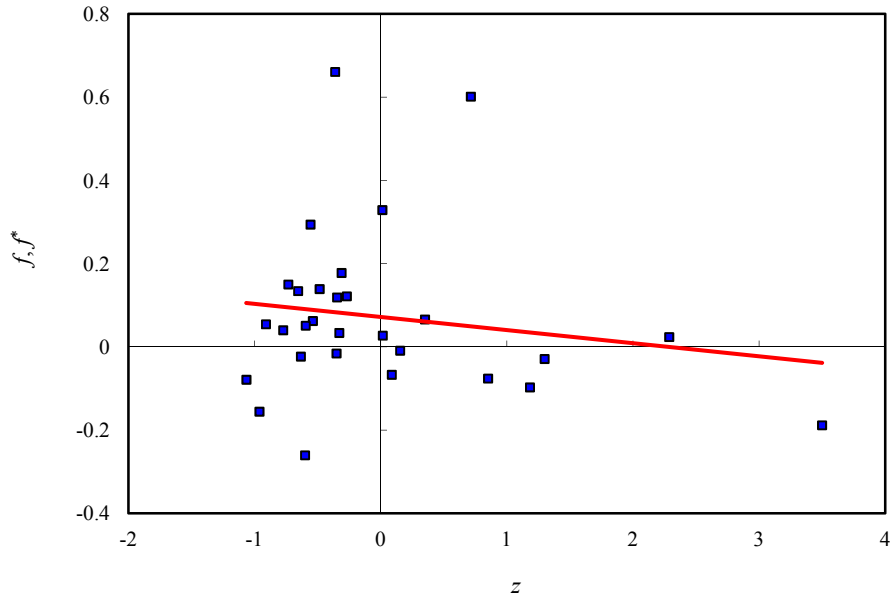

a. Based on inverse power function

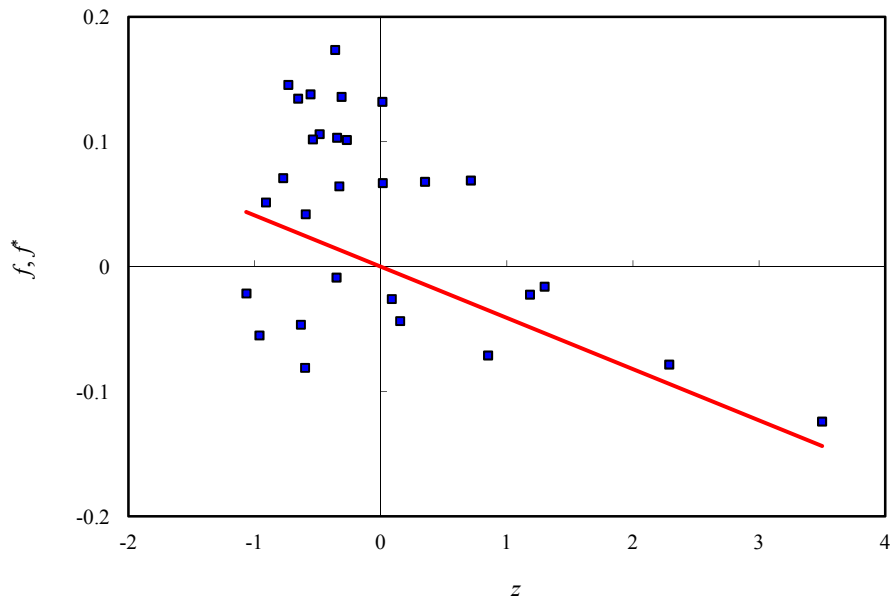

b. Based on negative exponential function

**Figure 3** The improved Moran's scatterplots with trendlines of spatial autocorrelation for the principal cities of China (2000)

Moran's scatterplots can be employed to make an analysis of spatial autocorrelation. It is easy to create Moran's scatterplots and the inverse Moran's scatterplots by means of the data in Table 3. Using $z$ to represent $x$-axis, and $f$ and $f^*$ to represent $y$-axis, we can draw the revised Moran's scatterplots, which are displayed in Figure 3. In the plot, the coordinates of observed values $(z, f)$ yield the scatter points, while the coordinates of expected values $(z, \hat{f}^*)$ give the trend line.

Accordingly, using $f^*$ to represent $x$-axis, and $z$ and $z^*$ to represent $y$-axis, we can generate the inverse Moran's scatterplots, which are shown in Figure 4. In this plot, the coordinates of expected values $(f^*, z^*)$ give the scatter points, while the coordinates of observed values $(f^*, z)$ yield the trend line. Obviously, the inverse Moran's plot is the mirror image of a Moran's plot. In other words, a Moran's scatterplot and its inverse scatterplot are reciprocal to one another. The reciprocal of the slope of the trend line in an inverse scatterplot equals the value of Moran's index.

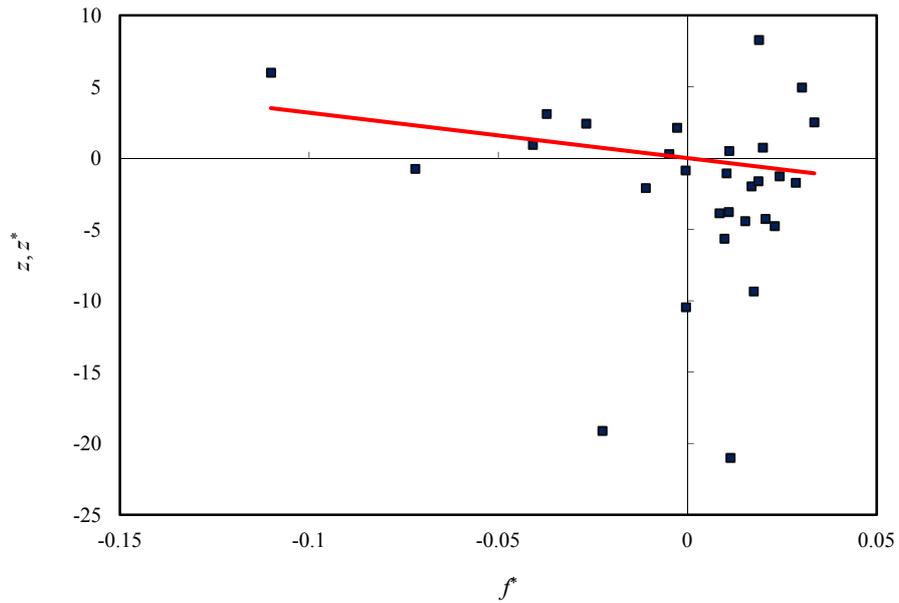

a. Based on inverse power function

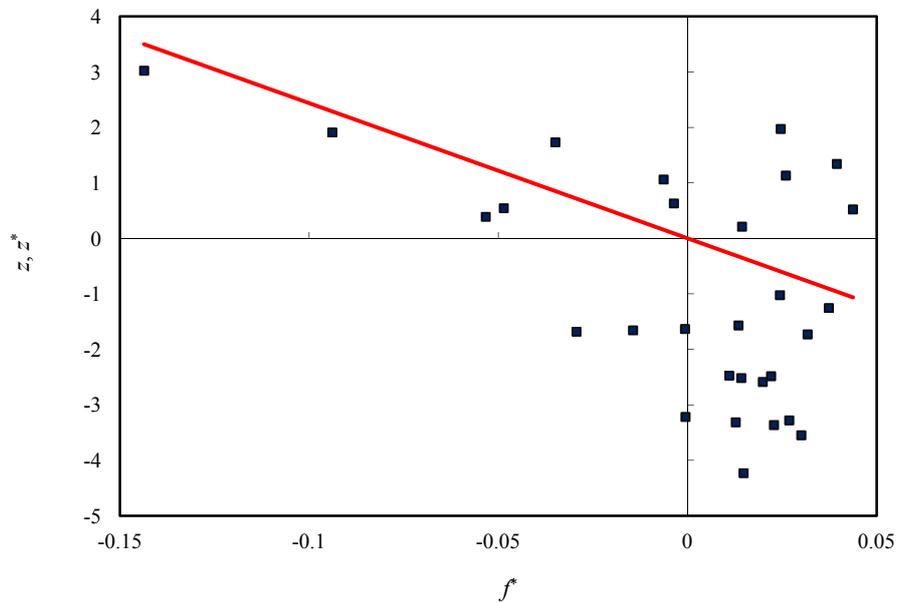

b. Based on negative exponential function

**Figure 4** The inverse Moran's scatterplots with trendlines of spatial autocorrelation for the principal cities of China (2000)

By employing the values of LISA and the revised or inverse Moran scatterplots, we can find the patterns of the spatial autocorrelation of China's cities in 2000. According to Moran's scatterplots, spatial autocorrelation falls into four types: the high-high correlation (H-H type: e.g. Tianjin, Shenyang, Harbin) in the first quadrant, the low-high correlation (L-H type: e.g. Hangzhou, Hefei, Nanchang) in the second quadrant, the low-low correlation (L-L type: e.g. Lanzhou, Xining, Yinchuan) in the third quadrant, and the high-low correlation (H-L type: e.g. Shanghai, Guangdong, Chongqing) in the fourth quadrant. The cluster result based on the power law is similar to that based on the exponential law; however, Beijing is an exception. In terms of the power function, Beijing belongs to the high-high type, but it falls into the high-low type in terms of the exponential function (Table 3). This indicates that, as a whole, spatial weight functions have no significantly different influence on the autocorrelation pattern of China's cities. Going a step further, we can find the most prominent Chinese cities through the process of spatial autocorrelation. Several evidences show that the city of Shanghai seems to be an exceptional case in Moran's scatterplots. First, the standardized sizes of two cities, Shanghai and Beijing, are greater than 2, the value of double standard deviation. Second, the maximum value of LISA belongs to Shanghai. Third, if we remove Shanghai from the sample, the trendline in Figure 3(a) will become nearly flat. These calculations show that the most prominent city of China in the spatial autocorrelation process is Shanghai followed by Beijing, Tianjin, and Hangzhou. This conclusion lends further support to the spatial correlation analysis of China's cities (Chen, 2009).

If we examine the error frequency distributions, we can determine the difference between the effect of the inverse power function and that of the negative exponential function. By the results displayed in Table 3, the residual values of spatial autocorrelation can be calculated using equation (33). Then, the bar graphs of frequency distributions based on the power function, equation (55), and the exponential function, equation (56), can be illustrated as follows (Figure 5). The graphs are expected to be bell-shaped histograms. However, both of the histograms fall short of expectations, but the second one, shown in Figure 5(b), is closer to the bell curve than the first one, shown in Figure 5(a). Based on the power function, the squared sum of the errors between the real frequency distribution and the theoretical normal distribution is about 0.1963, while based on the exponential function, the corresponding error sum of square is about 0.1487. This seems to suggest that the negative exponential function is more appropriate for the spatial autocorrelation analysis of China's cities than the inverse power function as a spatial weight function. As a matter of fact, the dataset consisting of only 29 elements is not large enough to form an unambiguous normal curve or bell histogram. The above-stated analytical process of normal histograms is just to demonstrate an application of a method, but the conclusion is for reference only.

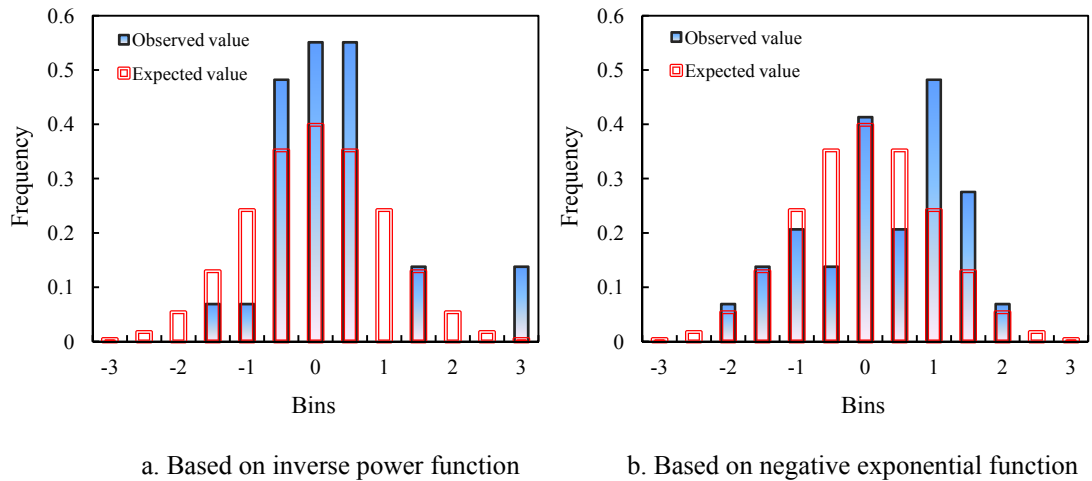

a. Based on inverse power function    b. Based on negative exponential function

**Figure 5** The normal histograms of error distributions based on different spatial weight functions for China's cities (2000) [**Note**: These graphs are created using standardized error series. The filled bars represent the actual distributions based on observed values, while the unfilled bars with double frames represent the normal distributions based on the expected values, which form bell-shaped histograms.]

In practice, the spatial population and the spatial sample are relative. Whether a dataset is treated as sample or population depends on the aim of spatial analysis. For the above example, the 29 cities in China can be regarded as a spatial population or a sample. If we investigate the capital cities only, the set of cities can be thought of as a population; but if we examine all the cities of China by means of this subset of cities, the 29 cities will make up a sample. Applying the approaches to evaluating Moran's index and Geary's coefficient to the dataset of the 29 China's cities yields a series of results of Moran's index and Geary's coefficient, which are tabulated below (Table 4). Clearly, the values of the different autocorrelation parameters based on different weight functions led to the same conclusion: weak negative spatial autocorrelation. This also suggests that if the size of dataset $n$ is large enough, the parameter values based on SSD are not significantly different from those based on PSD.

**Table 4** Moran's index and Geary's coefficient values based on spatial population and sample of Chinese cities (2000)

| Type | Parameter | Based on inverse power law | Based on negative exponential law | Threshold/expected value |
|---|---|---|---|---|
| For spatial population | Moran's $I$ | -0.0315 | -0.0410 | $0^a$ |
| | Geary's $C^*$ | 1.1543 | 1.0949 | $1^a$ |
| | Constant, $\omega$ | 1.1229 | 1.0538 | $1^b$ |
| For spatial sample | Moran's $I^*$ | -0.0304 | -0.0396 | $0^a$ |
| | Geary's $C$ | 1.1145 | 1.0571 | $28/29 \approx 0.9655^a$ |
| | Constant, $\psi$ | 1.0184 | 1.0175 | $28/29 \approx 0.9655^b$ |

**Note**: a—threshold value; b—expected value. The expected values of the constants are the threshold values of Geary's coefficient.

# 5 Conclusions

The significance of this work is that it provides a new approach to and a new way of understanding spatial autocorrelation analysis. In particular, based on the reformative expression of Moran's index, the spatial autocorrelation can be associated with scaling analysis. By mathematical transform, the relationship between Moran's index and the geographical scaling process can be revealed, and the results will be reported in an upcoming article. Now, three conclusions can be drawn from this study. **First, the spatial autocorrelation analysis can be simplified by means of matrix calculus**. Using equations of matrices, we can calculate Moran's index or Geary's coefficient by several steps with ease. If the global Moran's index is evaluated, then the local Moran's index can be incidentally obtained. What is more, based on the matrix expression, Moran's scatterplot can be improved and the inverse Moran's scatterplot can be put forward. **Second, the scopes of application of Moran's index and Geary's coefficient are different**. Moran's index is based on spatial population, while Geary's coefficient is based on spatial sampling results. If we plan to apply Moran's index to spatial samples, the formula of Moran's index should be revised; if we plan to apply Geary's coefficient to spatial population, the formula of Geary's coefficient should also be revised. **Third, the error frequency distribution can be employed to choose a spatial weight function**. One academic contribution of this paper to geographical analysis is the error formula of spatial correlation. More than one type of weight functions has been used to make spatial autocorrelation analysis; however, using the one resulting in an error frequency distribution which is more similar to Gaussian distribution is more advisable.

**Acknowledgement:**

This research was sponsored by the National Natural Science Foundation of China (Grant No. 41171129. See: http://isisn.nsfc.gov.cn/egrantweb/). The support is gratefully acknowledged. I would like to thank my student, Ms. Ruoxuan Xiong, for assistance in partial mathematical derivation of Moran's index. I am also grateful to two anonymous reviewers whose interesting comments were very helpful in improving the paper's quality.